\documentclass{article} 
\usepackage{glas}
\usepackage{psfig}  \usepackage{epsfig}
\def\lsim{\mathrel{\rlap{\lower4pt\hbox{\hskip1pt$\sim$}}
    \raise1pt\hbox{$<$}}}                
\def\gsim{\mathrel{\rlap{\lower4pt\hbox{\hskip1pt$\sim$}}
    \raise1pt\hbox{$>$}}}                

\begin{document}
\begin{titlepage}{GLAS-PPE/2000-01}{BRIS/HEP/2000--01}{March 2000}
\title{The Evolution of the
MLLA Parton Spectra.} 
\author{N.~H.~Brook\Instref{bris} I.~O.~Skillicorn\Instref{glas}}
\Instfoot{bris}{H. H. Wills Physics Lab., University of Bristol, UK.}
\Instfoot{glas}{Dept. of Physics \& Astronomy, University of Glasgow, UK.}
\begin{abstract}
The evolution with energy scale of the partonic logarithmic scaled energy
spectra is investigated in the framework of the modified 
leading logarithmic approximation (MLLA).
The behaviour of the higher order moments 
is compared to a number of analytic 
predictions and $\rm e^+e^-$ data.
\end{abstract}
\end{titlepage}

\section{Introduction}
Quantum Chromodynamics (QCD)
currently does not allow direct calculations of the
hadronic final state observables. To make predictions of the
final state it is necessary to
model the transition from partons to hadrons.
One approach that has been successful
in describing the general features of the inclusive energy spectra,
in both $\rm e^+e^-$ annihilation and deep inelastic
scattering experiments~\cite{MLLArev},  is the modified leading log
approximation (MLLA) using local parton hadron duality (LPHD) as the
method for relating the MLLA partonic predictions to the hadronic
observables.

The production of hadrons in hard scattering processes is controlled by the
underlying partonic behaviour. 
At small values of momentum fraction (of the outgoing
parton from the original hard
scatter) this parton
behaviour, often referred to as a parton shower, is
dominated by gluon bremsstrahlung. The branching processes $q
\rightarrow qg$ and $g \rightarrow gg$ (double logarithmic
processes) in addition to $g \rightarrow q \bar q$ (single logarithmic
process) give rise to this parton shower. MLLA accounts for both these
double
and single logarithmic effects in the evolution equations~\cite{DGLAP}.
The perturbative
properties of partonic distributions have been calculated in the framework of
the MLLA. They are
governed by two free parameters: a running strong coupling, defined by
a QCD scale $\Lambda,$ and an energy cut-off, $Q_0,$ below which the
parton evolution is truncated.

Using LPHD, the non-perturbative effects of particle
distributions are reduced to a simple factor of normalisation that
relates the hadronic distributions to the partonic ones.
At large enough energies, away from the influence of $Q_0,$
 this `hadronisation' factor should be 
independent of the energy scale~\cite{LPHD}
 at which the spectra are being calculated.


Various approaches, detailed below, have been taken to calculate
the single particle spectra and their moments within the MLLA framework.
In this paper these theoretical approaches are compared as a function of 
energy scale, $Q.$ Substantial difference are found in the 
predictions of the higher moments.
The theoretical results, in particular the higher order moments,
are also compared with $\rm e^+e^-$ data
over a range of centre of mass energies, $Q=E_{CM}/2.$ It is found that
the general characteristics are well described but that 
the theory is not in accord with all features of the data
if it is assummed that $\Lambda,\  Q_0$ and the LPHD normalisation 
are energy independent.
The studies
reported here avoid the need to extrapolate into regions not
experimentally measured.

\section{Single Particle Spectra}
Given a high energy parton which fragments via secondary partons
into a jet of hadrons,
the MLLA evolution equation allows the secondary parton spectra for the
logarithmic scaled energy, $\xi,$ to be calculated~\cite{dokevol}. 
The variable $\xi$ is
defined as $\ln(Q/E) \equiv \ln(1/x),$ 
where $Q$ is the energy of the original parton 
and $E$ is the energy of the secondary 
parton. The cut-off, $Q_0,$ limits the parton
energy to $ E \ge k_T \ge Q_0,$ where $k_T$ is the transverse energy of
the decay products in the jet evolution. In order
to reconstruct the $\xi$ distributions
an inverse Mellin transformation is performed

\begin{equation}
\frac{1}{N}\frac{dn_h}{d\xi} \propto
\bar D(\xi,Y,\lambda) = \int_{\epsilon - \imath\infty}^{\epsilon + \imath\infty}
\frac{d\omega}{2\pi \imath} x^{-\omega} D(\omega,Y,\lambda)
\label{eqn:MLLA}
\end{equation}
where the integral runs parallel to the imaginary axis on the right of
all singularities in the complex $\omega-$plane, $Y = \ln(Q/Q_0)$
and $\lambda=\ln(Q_0/\Lambda).$

The Mellin-transformed distributions, $D(\omega,Y,\lambda),$ can be
expressed~\cite{dokevol}  
in terms of confluent hypergeometric functions, $\Phi,\ $ as

\begin{equation}
\begin{array}{lc}
D(\omega,Y,\lambda) &=  \frac{t_1A}{B(B+1)}\Phi(-A+B+1,B+2;-t_1)\Phi(A-B,1-B;t_2) \\
 &+\left(\frac{t_2}{t_1}\right)^B\Phi(-A,-B;-t_1)\Phi(A,B+1;t_2),
\end{array}
\end{equation}
where 
\begin{equation}
\begin{array}{lr}
t_1 = \omega(Y+\lambda) \;\;,&  \;\; t_2= \omega\lambda
\end{array}
\end{equation}
\noindent and $A$ and $B$ are defined as

\begin{equation}
\begin{array}{lr}
A=4N_c/b\omega,\;\; & \;\;B=a/b,
\end{array}
\end{equation}
where $N_c$ is the number of colours, $a=11N_c/3+2n_f/3N_c^2$ , $n_f$
is the number of flavours and $b = 11N_c/3 + 2n_f/3.$
Throughout this paper the number of flavours is assumed to be 3.
Equation~(\ref{eqn:MLLA}) is then calculated using a numerical integration
in the complex $\omega-$plane.

It is convenient to investigate the MLLA spectra in terms of moments.
 The cumulant moments of the $\xi$ distribution can be written as:

\begin{equation}
K_m =
\left. \left(-\frac{\partial}{\partial\omega}\right)^m D(\omega,Y,\lambda)
\right|_{\omega=0}.
\end{equation}

\noindent The analytic form of these cumulant moments have been calculated 
in~\cite{dokevol} for the first four moments. This allows the
normalised moment, $\langle \xi^m \rangle,$ to be calculated and hence
the dispersion ($\sigma$), skewness ($s$) and the kurtosis ($k$) 
to be constructed, see for example ~\cite{kendall} 

\begin{eqnarray}
\langle \xi \rangle & = & K_1  \\
\nonumber \\
\sigma^2 & = & K_2  = \langle\xi^2\rangle - \langle\xi\rangle^2 \\ 
\nonumber \\
s & = & K_3/\sigma^3  = 
\frac{\langle\xi^3\rangle-3\langle\xi^2\rangle\langle\xi\rangle+2\langle\xi
\rangle^3}{\sigma^3} \\ 
\nonumber \\
k & = & K_4/\sigma^4  = 
\frac{\langle\xi^4\rangle-4\langle\xi^3\rangle\langle\xi\rangle-
3\langle\xi^2\rangle^2+12\langle\xi^2\rangle\langle\xi\rangle^2
-6\langle\xi\rangle^4}
{\sigma^4}.
\end{eqnarray}

The so-called limiting spectrum is the case when $Q_0=\Lambda$ i.e.
$\lambda=0$.  For this case, Fong and Webber~\cite{fongweb} have
also calculated the behaviour of the moments of the $\xi$
spectra with energy scale, $Q.$
They point out that the spectra can be represented close to the
maximum of the distribution  
by a distorted Gaussian of the form:

\begin{equation}
\bar D(\xi,Y) \propto \exp\left[ \frac{1}{8}k-\frac{1}{2}s\delta
-\frac{1}{4}(2+k)\delta^2 + \frac{1}{6}s\delta^3
+\frac{1}{24}k\delta^4 \right]
\label{eq:distg}
\end{equation}
where  $\delta = (\xi - l)/\sigma$ and $l$
is the mean. 
This expression, when fitted to the spectrum, 
allows the moments to be determined
when the full spectrum is either unmeasured or uncalculable.

The changes to the parton spectra for quark, $q,$ and gluon, $g,$ jets
are Next-to-MLLA effects which result in additional terms in the
integral shown in Eq.~(\ref{eqn:MLLA}).
These corrected spectra (at $Q_0 = \Lambda$) for the quark and gluon jet
can be related to the limiting spectra~\cite{pQCD} as follows

\begin{equation}
\bar D_{q,g} = \left[ 1 +\Delta_{q,g}\left(\frac{\partial}{\partial
l}+\frac{\partial}{\partial Y}\right)\right]\bar D^{lim}(l,Y) \approx
\bar D^{lim}(l+\Delta_{q,g},Y+\Delta_{q,g}),
\label{eqn:qgspectra}
\end{equation}
\noindent with
\[
\Delta_g   =   -\frac{1}{3}n_f\frac{N_c^2-1}{2N_c^3}  \ ,
\ \Delta_q  =  \Delta_0 + \Delta_g \ ,
\ \Delta_0  =  \frac{a-3N_c}{4N_c}. 
\]

The effect on the moments is that the $\sigma,$
skewness and kurtosis for the quark/gluon distribution would be 
approximately the same as the limiting distribution at an effective
energy $Y+\Delta_{q,g}.$ The mean is the same as that of the
limiting spectra at this effective energy  but
shifted in value by $-\Delta_{q,g}.$
These shifts are small with the limiting spectra being between the
distribution for the quark and gluon jets.
It should be noted that the predictions of Fong and Webber~\cite{fongweb}
for the relative shifts between quark and gluon jets are slightly different
from those discussed above.

In the MLLA approach, the partons are assumed massless so the scaled energy 
and momentum spectra are identical. Experimentally the scaled momentum
distribution is usually measured and as the observed hadrons are massive the
equivalence of the two spectra no longer holds. In~\cite{klo} the assumption
is made that the cut-off $Q_0$ can be related to the masses of hadrons.
This allows the logarithmic
scaled momentum distribution, $\xi_p,$ to be written as

\begin{equation}
\frac{1}{N}\frac{dn_h}{d\xi_p} \propto
\frac{p_h}{E_h}\bar D(\xi,Y), 
\label{eq:xip}
\end{equation}
where 
$$
\xi = \log\frac{Q}{\sqrt{Q^2e^{-2\xi_p}+Q^2_0}},
$$
\noindent the energy of a hadron with a momentum $p_h$ is 
$E_h = \sqrt{p_h^2+Q_0^2}.$
Limiting momentum spectra based on massless partons and massive partons will
be referred to as MLLA-0 and MLLA-M spectra, respectively.

\section{Behaviour of Theoretical Spectra}
In this section the general characteristics of the spectra are discussed
in order to illustrate their behaviour as a function of the three 
variables $Q$, $\lambda$, and $Q_0$.

\begin{figure}[ph!]
\begin{center}
\mbox{\epsfig{file=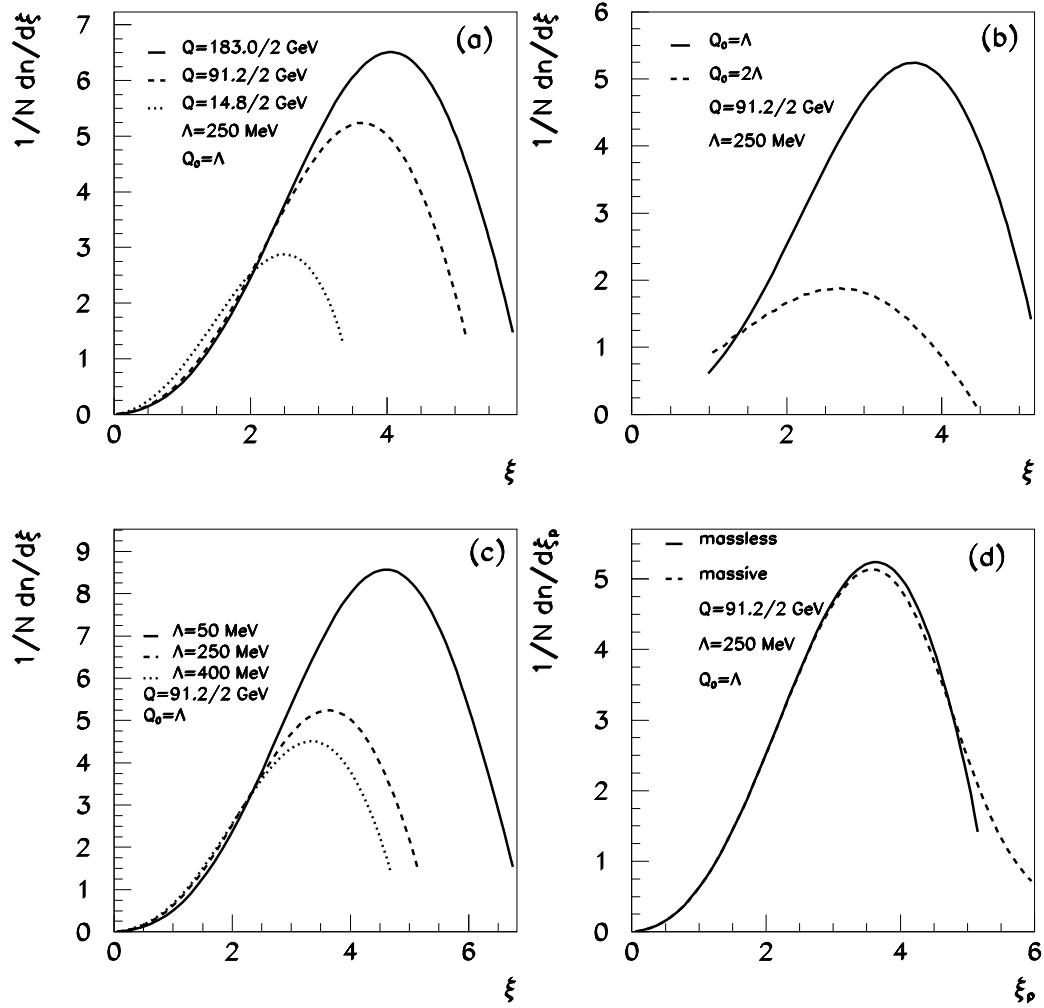,width=0.90\textwidth}}
\end{center}
\caption{ The logarithmic scaled energy (a)-(c) and
scaled momentum (d) spectra of partons as calculated
in the MLLA formalism. Unless otherwise stated values of $\Lambda=250{\rm\
MeV}$ and $Q_0=\Lambda$ (the so-called limiting spectra) at $Q=91.2/2{\rm\
GeV}$ were used to calculate the
curves.
(a) The three curves are predictions at three different energy scales,
$Q=183.0/2,91.2/2 {\rm\ and\ } 14.8/2 {\rm\ GeV}$ represented by the full,
dashed and dotted lines respectively. 
(b) The two curves are predictions for two different values of
$Q_0= \Lambda {\rm\ and\ } 2\Lambda$ represented by the full
and dashed lines respectively. 
(c) The three curves are predictions for three different values of
$\Lambda=50, 250{\rm\ and\ } 400 {\rm\ MeV}$ represented by the full,
dashed and dotted lines respectively. 
(d)The logarithmic
scaled momentum spectra of partons as calculated in the MLLA formalism.
The two curves are predictions for a massless (full line)
and massive assumption (dashed line).}
\label{xi}
\end{figure}

Figure~\ref{xi}(a) shows the limiting energy  
spectra for three energies
$Q=183./2$ (LEP II),  $91.2/2$ (LEP) and $14.8/2 {\rm\ GeV}$ (TASSO)
with a $\Lambda$ value of
$250{\rm\ MeV}.$ As the energy scale
increases, the parton multiplicity (the
area under the curve) grows and the peak position shifts to the right
i.e. the parton scaled energy spectra is softer.
The cut-off on the right hand side of the plot corresponds to the
truncation in the parton energy spectra at $Q_0 = \Lambda.$

Figure~\ref{xi}(b) shows the scaled energy spectra at fixed energy scale
$Q=91.2{\rm\ GeV}$ and  $\Lambda=250{\rm\ MeV}$ but at two different
values for the cut-off of the parton evolution, $Q_0 = \Lambda {\rm\ and\ }
2\Lambda.$ (The curves are cropped at $\xi=1.0$ as below this value
the numerical integration of the truncated spectra is unstable.)
Truncating the cascade at higher values of $Q_0$ leads to a
lower parton multiplicity with a harder energy spectra.

Figure~\ref{xi}(c), again, shows the limiting energy spectra but   
at a fixed energy scale, $Q=91.2/2{\rm\ GeV},$ with values of $\Lambda
= 50, 250 {\rm\ and\ } 400 {\rm\ MeV.}$ 
As $\Lambda$ decreases,
more partons are produced and their energy spectra is
softer. 
In the case of the limiting spectra,
this behaviour is dominated by the fact 
that an increase in $\Lambda$ is accompanied by an increase in the
cut-off $Q_0.$
 
Figure~\ref{xi}(d) shows the limiting scaled momentum spectra for the
MLLA-0 $(E_h=p_h)$ and MLLA-M case $(E_h \ne p_h)$ 
at an $E_{CM}=91.2 {\rm\ GeV}.$ The MLLA-0 
spectra displays the usual truncation at large $\xi$ associated with the
cut-off $Q_0,$ but the MLLA-M case
does not. This is due to the fact the calculation is now regulated by
the cut-off entering as a mass term in the expression of $\xi_p$ (see
Eq.~\ref{eq:xip}.) As the momentum $\rightarrow Q,$ i.e.
$\xi_p\rightarrow 0.0,$ the introduction of a mass term has no effect on
the MLLA calculation. As $\xi_p$ increases (the momentum becomes smaller)
the mass term begins to play an increasingly important r\^{o}le.
From these arguments it can be seen that as $Q$ decreases
mass term has a more significant influence over a larger (fractional)
range of the $\xi_p$ spectra.

\begin{figure}[ph!]
\begin{center}
\mbox{\epsfig{file=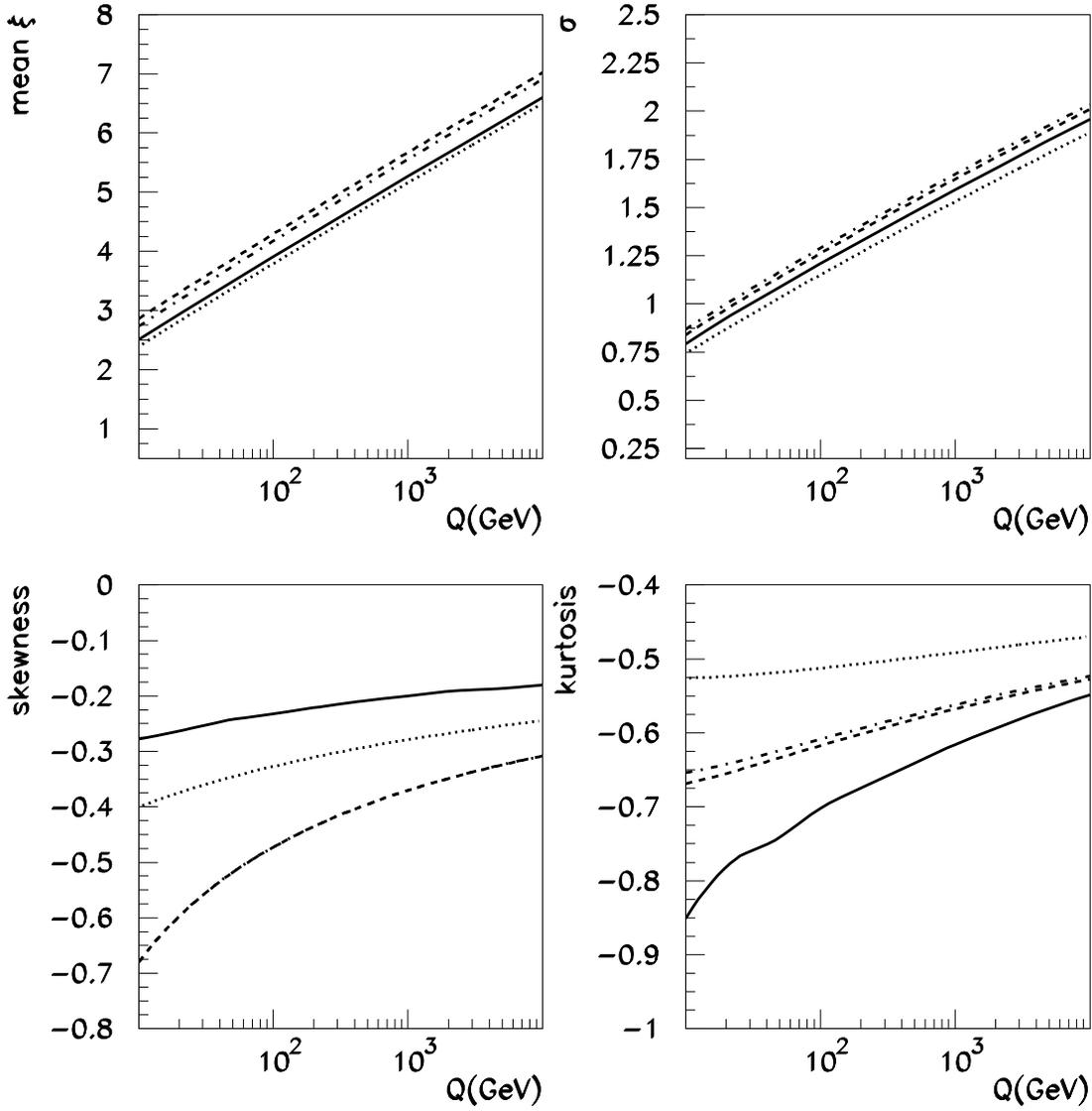,width=\textwidth}}
\end{center}
\caption{ The evolution of the mean, $\sigma,$ skewness and kurtosis as
a function of $Q$ for $\Lambda=250{\rm\ MeV.}$ The dashed and 
dash-dotted lines are the analytic 
calculation of Fong and Webber for the gluon and quark respectively, 
the dotted line from 
the analytic calculation of
Dokshitzer {\it et al.}, and the full line from fitting a distorted
Gaussian around $\pm1 \sigma$ of the mean of the MLLA-0 spectra.}
\label{evolve}
\end{figure}

Figure~\ref{evolve} shows the evolution of the mean, $\sigma,$ skewness and 
kurtosis of the limiting energy spectra with $\Lambda=250 {\rm\ MeV}$
as a function of $Q.$ Three different approaches to calculating the moments 
have been investigated: 
(i) the analytic expression of Fong and Webber for gluon jets (dashed line)
and quark jets(dash-dotted line); (ii) the 
analytic calculation of Dokshitzer {\it et al.}~\cite{dokevol} (dotted line); 
(iii) fitting a distorted
Gaussian (Eq.~\ref{eq:distg}) over $\pm 1\sigma$ around the mean value
(full line). The range of the fit is motivated by the 
phenomenological~\cite{fongweb} scope of validity.
All approaches exhibit the same trends, in that all the cumulants increase as
$Q$ increase. The values of skewness and kurtosis are negative, tending towards
zero as $Q$ increases, i.e. the spectra are becoming more like a pure Gaussian.
The mean and the $\sigma$ exhibit a very similar dependence on $Q$
for all three approaches, though with different offsets. The skewness and 
kurtosis exhibit a different dependence on $Q,$ but as $Q$ approaches
the asymptotic limit the predictions are converging. 

The differences between quark and gluon jets for
the predictions of Fong and Webber, in Fig.~\ref{evolve}, are small.
A feature of the data is the marked difference between the analytic calculations
of Dokshitzer {\it et al.}~\cite{dokevol}  and the fit to the limiting spectra. 
Calculating the moments over the full range of the spectra 
gives the same result as the analytic calculations, not too
surprisingly since both calculations use the same assumptions. 
Investigating a limited range of the spectra, as is done during the fit, 
produces marked differences,
highlighting the sensitivities of these higher-order moments to the 
behaviour of the tails as well as the incorrect form of the distorted Gaussian
away from the peak position. 
The difference between the analytic calculation of
Fong and Webber and that of Dokshitzer et al. could well be due to the 
differences in the predictions at large
$x_p,$ where the theoretical assumption made in both calculations are no
longer valid. 

Figures~\ref{xi} and~\ref{evolve} imply that $\Lambda$ 
influences the position of the maximum 
and the width of the distribution  
of $\xi$. The effect of the relation between $Q_0$ and $\Lambda$
is more complex; setting $Q_0 = 2\Lambda$ lowers the position of the 
maximum for a given $\Lambda$. The maximum can be returned to the
original position by reducing $\Lambda$ but with a consequential increase
of the width of the distribution relative to that
found for the limiting spectrum. The effect of introducing mass terms 
into the momentum spectra is to increase the tail of the spectra at high
$\xi,$ thus increasing skewness. As $Q$ increases, the influence of the
mass term becomes less important.

\section{Comparison with data}
\begin{figure}[ph!]
\begin{center}
\mbox{\epsfig{file=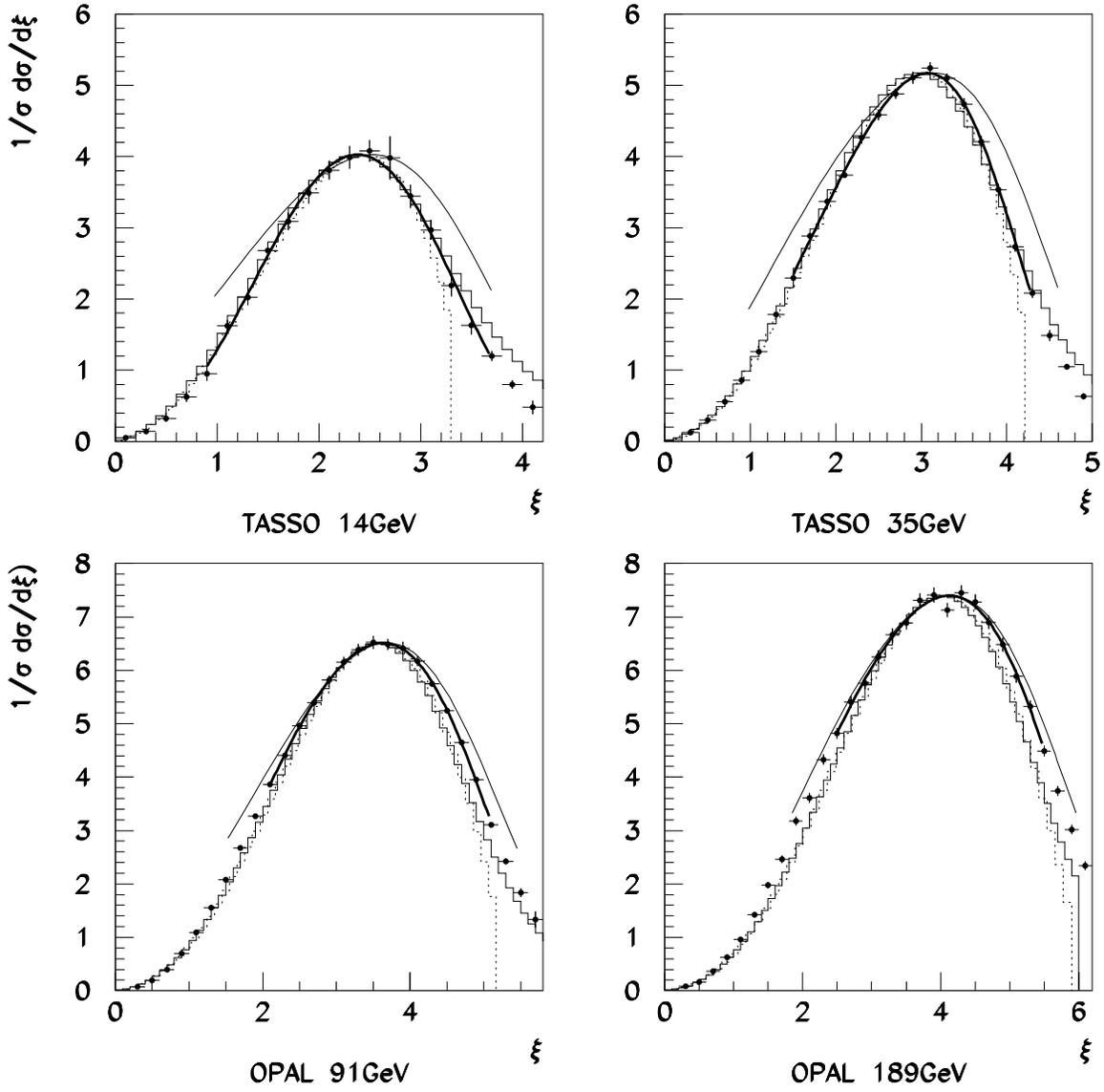,width=\textwidth}}
\end{center}
\caption{ $\rm e^+e^-$ data compared  with 
MLLA-0 spectra
( dashed histogram), MLLA-M spectra(
solid histogram ) and MLLA $Q_0 = 2\Lambda$ with 
$\Lambda = 50 {\rm\ MeV}$ ( thin solid  line).
The thick solid line is from  the fit of a distorted Gaussian   that
is used to determine the moments for the data.}
\label{eee}
\end{figure}

The differential $\xi_p$ distributions for $\rm e^+e^-$   data and the models
 are shown in Figure ~\ref{eee}.
The limiting spectra were calculated
with  $\Lambda=250{\rm\ MeV}$; this value was chosen to give the
correct peak position of the spectra for an $E_{CM}(=2Q)$ at the mass
of the $\rm{Z^0},\ M_Z.$ 
For each energy, 
the theoretical spectra are normalised to the same maximum height as the data.

The theoretical spectra
follow the general trend of the data for both  for both MLLA-0 and 
MLLA-M spectra;
in detail there are discrepancies. At low energies, for the MLLA-M case, the  
skewness is larger (i.e. more positive ) than that observed for the data;  
at high energies, the limiting spectra have a 
smaller width than the data. 
Also shown are the MLLA energy spectra for $Q_0 = 2\Lambda$. Here 
$\Lambda = 50 {\rm\ MeV}$  is required
to get the correct position of the maximum for 
$E_{CM}=\rm M_Z.$ The position of the maximum 
has approximately the correct energy dependence; 
however, the width of the theoretical
spectrum is consistently greater than that of the data.

To study the energy evolution of the MLLA  theoretical spectra
and the $\rm e^+e^-$ data, fits around the peak
position, using the distorted Gaussian (Eq.~\ref{eq:distg})
have been  made to determine the moments of the $\xi_p$ distributions.
In view
of the statistical limitations, the range of the fit for the data
was about three units around the peak
(see  Figure~\ref{eee}). The distorted Gaussian describes the data well
over the fitted region with a $\chi^2/dof$ of 1.0 or better.
  To permit direct comparison with the data,
a similar fit range was used for the theoretical spectra.
The distorted Gaussian  gives a good description of  the 
MLLA-M  spectra. The description is less good for 
the MLLA-0 spectra but for all energies the Gaussian represents
the model to better than $1\%$.
 
\begin{figure}[ph!]
\begin{center}
\mbox{\epsfig{file=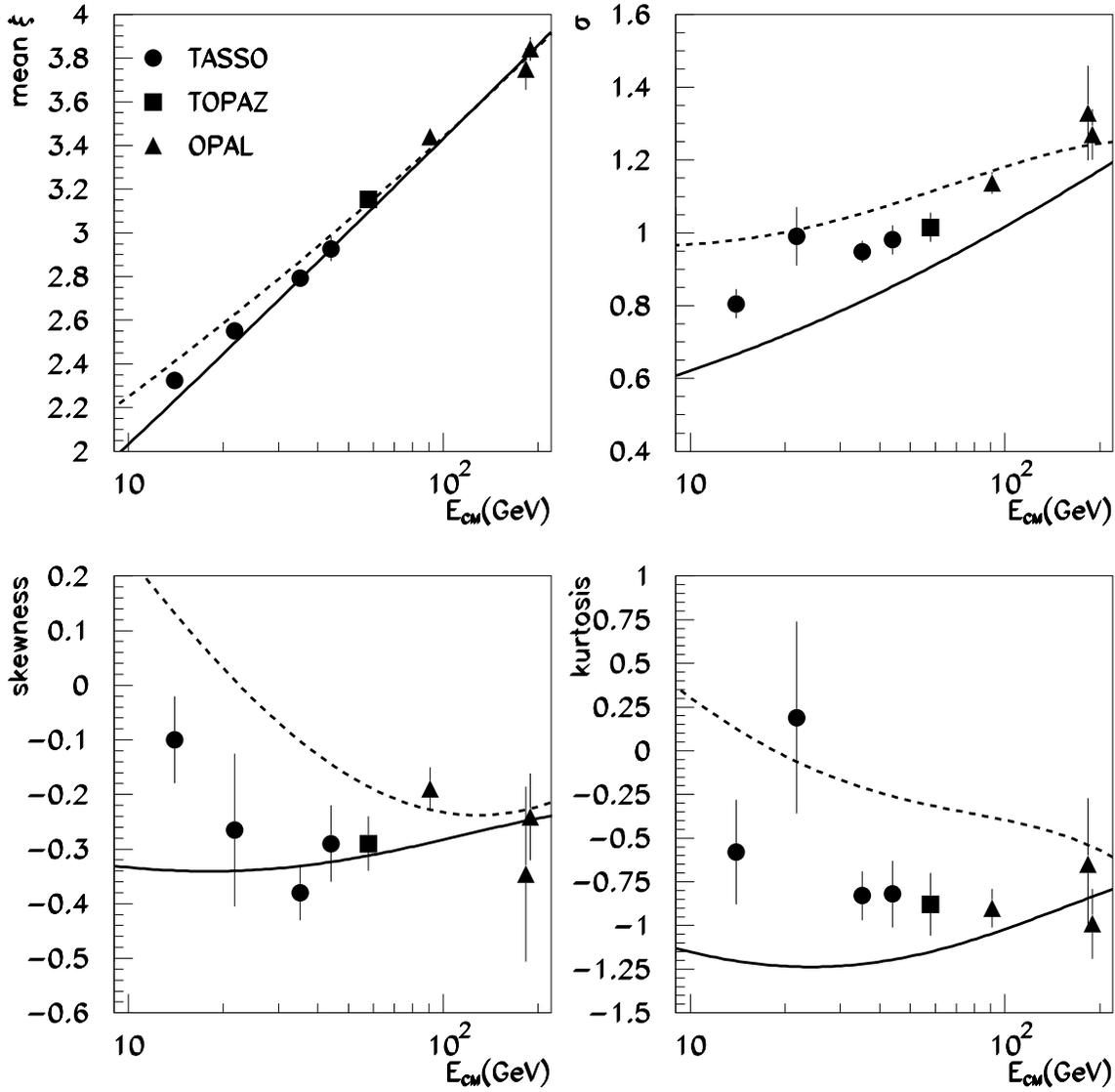,width=\textwidth}}
\end{center}
\caption{Moments as a function of $E_{CM}(=2Q).$
The   points with error bars are from fits of a 
distorted Gaussian to $\rm e^+e^-$ data;
circles: TASSO;
square: TOPAZ; triangles: OPAL.
The solid lines are moments  from a fit of a distorted Gaussian to the
MLLA-0 spectrum. The dashed lines are moments from a fit of
a distorted Gaussian to a  
MLLA-M spectrum. }
\label{skew}
\end{figure}

The moments from the fits  are shown in Figure~\ref{skew}
as  solid
lines for fits to MLLA-0 spectra and as dashed lines
for fits to the MLLA-M spectra. The data points are from our fit to 
$\rm e^+e^-$ data~\cite{eedata,OPAL}.

It may be concluded from Figure~\ref{skew} that the MLLA-0 
model gives a good description of the $\rm e^+e^-$ data for 
the mean and skewness.
There are however discrepancies in $\sigma$ and kurtosis as 
is also  evident in  Figure~\ref{eee}; $\sigma$ is smaller, 
and the model is more platykurtic than  the data. 
The  MLLA-M model  gives a poorer description of
all variables at low $E_{CM}$ but approaches both the MLLA-0 predictions  and
the data at high  $E_{CM}$.

This is contrary to the conclusions reported by Lupia and 
Ochs~\cite{lupochs}. In that
paper the moments of the experimental distributions corrected for the mass
effects were compared to the predictions for the limiting spectra and
found to be in good agreement. It
should be noted that in~\cite{lupochs} an extrapolation was made
into regions where no experimental measurement exists to allow the
moments to be calculated.
In addition, the agreement of the MLLA limiting spectra
with the mass corrected experimental $\xi$ distribution is poor. 
The spectra at low $\xi$ 
are well described but the region around the peak and at higher $\xi$ is
generally not well reproduced. 
The studies
reported here avoid the need to extrapolate into regions not
experimentally measured.

\begin{figure}[ph!]
\begin{center}
\mbox{\epsfig{file=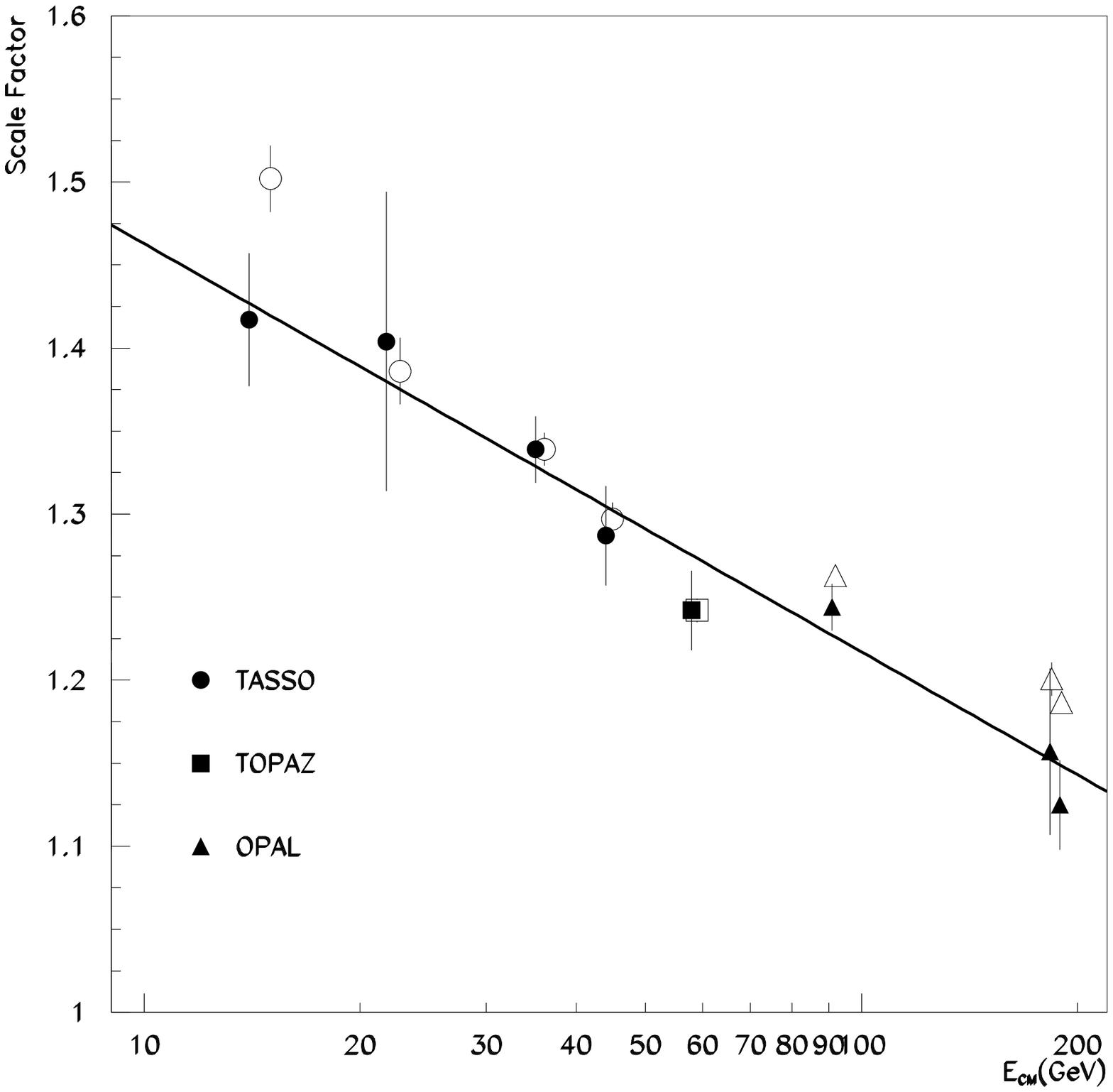,width=\textwidth}}
\end{center}
\caption{Scale factor as a function of $E_{CM}(=2Q).$
The full symbols are from normalisation to the same maximum height
for data and MLLA-0  spectra.
The open symbols (displaced by 1 GeV for clarity) are from a 
normalisation of the areas of data
and MLLA-M spectra;
circles: TASSO;
square: TOPAZ; triangles: OPAL.
The line is  a linear fit of the height-based scale factor to $\log(E_{CM})$.}
\label{LPHD}
\end{figure}

LPHD postulates that a constant factor relates the MLLA predicted spectrum
to the experimental spectrum, independent of the $E_{CM}.$
In the spectra discussed above,
the maxima of the MLLA predictions have been scaled to the data.
This scale factor, for the MLLA-0 model,
is shown in  Figure~\ref{LPHD} as a function of  $E_{CM}.$
It falls from $\approx 1.4$ at the lowest $E_{CM}$ to $\approx
1.2$ at the highest.
A similar result is obtained for an area normalisation over approximately
three units of $\xi_p$ for the MLLA-M case. Thus our observations
disagree with the LPHD postulate of a constant scale factor.
This suggests that at the currently accessible experimental energies the
theoretical spectra are still influenced by the $Q_0$ cut-off.

Similar comparisons of $\rm e^+e^-$ data with MLLA predictions
have been made by the DEPHI~\cite{DELPHI} and OPAL~\cite{OPAL}
collaborations. CDF has  examined jets produced in $p\overline{p}$
interactions~\cite{CDF}. In contrast to the analysis presented here, 
no explicit analysis of the
higher moments of the $\xi$ distribution has been made.
However, where the analyses can be compared, the results are in agreement
with those presented here. In particular, $\Lambda$ is close to  
250MeV, the shape of the MLLA predicted spectrum is close to that
of the data but disagrees in detail, and the value of the LPHD constant
changes with energy.


\section{Conclusions}
The MLLA predictions of the logarithmic scaled energy spectra, and in
particular their moments, have been investigated as function of energy
scale.
Various theoretical approaches have been compared and
differences discussed. We note that care should be taken in the comparisons 
to ensure that a consistent approach is maintained; in particular attention
should be paid to the range of application in $\xi.$
There is a large discrepancy in the skewness and kurtosis at low $Q$ but
the various predictions  converge as the asymptotic limit is
approached. 

The theoretical results are compared to measurements taken in $\rm
e^+e^-$ annihilation experiments. The calculations are generally in good
agreement with the data. It is observed that the limiting spectra is
preferred over the predictions of the truncated cascade ($Q_0 \ne
\Lambda$.)
The introduction of the mass term has a large effect at all but the
highest  $E_{CM}.$ The data is consistently broader than
the  limiting spectra over the energy range studied here.

The normalisation factor of LPHD between the data and the MLLA
theoretical predictions is not constant. Contrary to expectation
it displays a
dependence on $E_{CM},$ decreasing as
$E_{CM}$ increases, thus suggesting a residual influence of the $Q_0$
cut-off.

\section*{Acknowledgements}
The authors would like to thank Yuri Dokshitzer for useful discussion and
clarification of issues.

\newpage

\newpage

\newpage

\newpage

\newpage


\begin{thebibliography}{99}

\bibitem{MLLArev}
V.A.~Khoze and W.~Ochs, Int. J. Mod. Phys.  A12 (1997) 2949.

\bibitem{DGLAP}
V.N.~Gribov and L.N.~Lipatov,
\newblock  Sov.\ J.\ Nucl.\ Phys.\ 15 (1972) 438 and 675;\\
\newblock Yu.L.~Dokshitzer, Sov.\ Phys.\ JETP\  46 (1977) 641;\\
G.~Altarelli and G.~Parisi,  Nucl.\ Phys.\  B126 (1977) 298.

\bibitem{LPHD} Ya. I. Azimov et al., Z. Phys. C27 (1985) 65 and C31
(1986) 213.

\bibitem{dokevol} Yu. Dokshitzer, V.~Khoze and S.~Troyan, Int. J. Mod.
Phys.  A7 (1992) 1875.

\bibitem{kendall} A.~Stuart and J. K. Ord, `Kendall's Advanced Theory of
Statistics', Griffin (1987).

\bibitem{fongweb} C.\ P.\ Fong and B.~R.~Webber, Phys. Lett. B229 (1989)
289;\\
C.\ P.\ Fong and B.~R.~Webber, Nucl. Phys. B355 (1991) 54.
                                                             
\bibitem{pQCD} Yu. Dokshitzer et al., `Basics of Perturbative QCD',
Editions Fronti\`eres (1991).

\bibitem{klo} V.~Khoze, S.~Lupia and W.~Ochs, Phys.\ Lett.\  B386 (1996) 451.

\bibitem{eedata} 
OPAL Collab., R. Akers et al., Z. Phys C68 (1995) 203; \newline
OPAL Collab., G. Alexander et al., Z. Phys. C72 (1996) 191; \newline
OPAL Collab., K. Ackerstaff et al., Z. Phys. C75 (1997) 193; \newline
TASSO Collab., W.~Braunschweig et al., Z. Phys. C47 (1990) 187; \newline
TOPAZ Collab., R.~Itoh et al., Phys. Lett. B345 (1995) 335.

\bibitem{OPAL} OPAL Collab., CERN-EP/99-178, hep-ex/0002012.

\bibitem{lupochs} S.~Lupia and W.~Ochs, Eur.\ Phys.\  J.\ C2 (1998) 307.

\bibitem{DELPHI} DELPHI Collab.,CERN-EP/99-57

\bibitem{CDF} CDF Collab., EPS-HEP99 paper 1\_600.
 
\end{thebibliography}
\end{document}